# Ultra-compact Si/In$_2$O$_3$ hybrid plasmonic waveguide modulator with a high bandwidth beyond 40 GHz

YISHU HUANG,[1] JUN ZHENG,[1] BINGCHENG PAN,[1] LIJIA SONG,[1] GUANAN CHEN,[1] ZEJIE YU,[1] HUI YE,[1] AND DAOXIN DAI[1,2,*]

[1] State Key Laboratory for Modern Optical Instrumentation, Center for Optical & Electromagnetic Research, College of Optical Science and Engineering, International Research Center for Advanced Photonics, Zhejiang University, Zijingang Campus, Hangzhou 310058, China.
[2] Ningbo Research Institute, Zhejiang University, Ningbo 315100, China.
*dxdai@zju.edu.cn

**Abstract:** Optical modulators are required to have high modulation bandwidths and a compact footprint. In this paper we experimentally demonstrate a novel Si/In$_2$O$_3$ hybrid plasmonic waveguide modulator, which is realized by an asymmetric directional coupler (ADC) consisting of a silicon photonic waveguide and a Si/In$_2$O$_3$ hybrid plasmonic waveguide. The optical signal is modulated by radio-frequency (RF) signal applied on the Au electrodes at the top of MOS capacitor and contacting the In$_2$O$_3$ thin film. The record-high modulation bandwidth of >40 GHz is realized by a silicon-doping-free metal-oxide-In$_2$O$_3$ capacitor integrated in a 3.5-μm-long asymmetric directional coupler (ADC) for the first time.

## 1. Introduction

Optical interconnect systems increasingly desire optical modulators with high modulation bandwidth and compact footprint to transmit huge data with dense devices integrated on a single chip [1]. Toward this goal, people have proposed and demonstrated modulators based on doped silicon microring [2], hybrid silicon/III-V-semiconductor cavity [3], lithium niobate (LN) Mach-Zehnder interferometer (MZI) [4,5] and plasmonic structures with organic electro-optic (OEO) materials [6,7]. Recently, transparent conductive oxides (TCOs), with its strong plasma dispersion and epsilon-near-zero (ENZ) effects [8–14], gradually emerge as an alternative choice for on-chip optical modulation. Firstly, TCO-based electro-absorption [15] and nanocavity modulators [16] have advantages on its microns-level device length for higher packing density. Then, the strong modulation strength in a small footprint brings low driving voltage as one volt [17] and <100-fJ/bit energy consumptions [17–19]. When applied with microring resonators, the TCO-based ones exhibit a wavelength tenability of 271 pm/V [20]. The MZI modulators with a sub-wavelength-long TCO-based active region have an ultra-low $V_\pi L$ of 95 V·μm [21].

Given the strong static modulation strength, people tend to investigate the high-speed performance of TCO-based modulators. Due to the carrier transportation time in TCOs is 200~300 ps [22], the main limitation of modulation bandwidth is the resistance-capacitance (RC) delay. All TCO-based modulators apply TCOs with a metal-oxide-semiconductor (MOS) capacitor [8,9]. For the TCO-oxide-silicon MOS capacitors, the large resistance of silicon, which is high as >1 MΩ, causes that the RC bandwidth is below GHz [16,17]. Then the silicon doping method is introduced to lower the silicon resistance and the modulation bandwidth is elevated to 1.94 GHz [18]. Apart from that, the metal-TCO-oxide-silicon MOS capacitor, which can support a plasmonic mode with sub-wavelength field confinement, reduces the series resistance due to the highly-conductive metal cap. Assisted with doping in the top layer of silicon, the electro-absorption modulators (EAMs) show modulation bandwidths of 1.25-GHz [23] and 3.5-GHz [24], respectively. Recently, the metal-oxide-TCO MOS capacitor, which totally gets rid of doped silicon in its electrical characteristics, is applied as an active region in a MZI and shows a 1.1-GHz modulation bandwidth [21]. According to the reports so far, it can be seen that various schemes of MOS capacitors cannot break the modulation limitation of GHz, which is far behind that of commercial doped-silicon modulators. The modulation bandwidth potential remains a huge problem for development of TCO-based modulators. Furthermore, the MOS capacitance varies with the conductivity of thin-film TCO in a complex way [24], which makes the experimental results of breakthrough on modulation bandwidth highly desired.

In this work, we experimentally demonstrate a plasmonic modulator based on hybrid integration of silicon and indium oxide ($In_2O_3$) in an asymmetric directional coupler (ADC). The footprint reducing of an ADC under certain phase matching condition makes that the MOS area (capacitance relevant) is quadratically decreased and device length (resistance relevant) is linearly increased. The more rapid variation of MOS area brings a decreased RC time constant. Different from the MOS capacitors in former designs, the metal-oxide-$In_2O_3$ capacitor in the ADC can be an extremely narrow one with no loss from mode mismatch and the resistance is lowered by the silicon-doping-free design. When the device length is reduced to 3.5 μm, a record-high >40 GHz modulation bandwidth (setup limited). This device also has an 80-nm optical bandwidth (1530–1610 nm).

## 2. Design

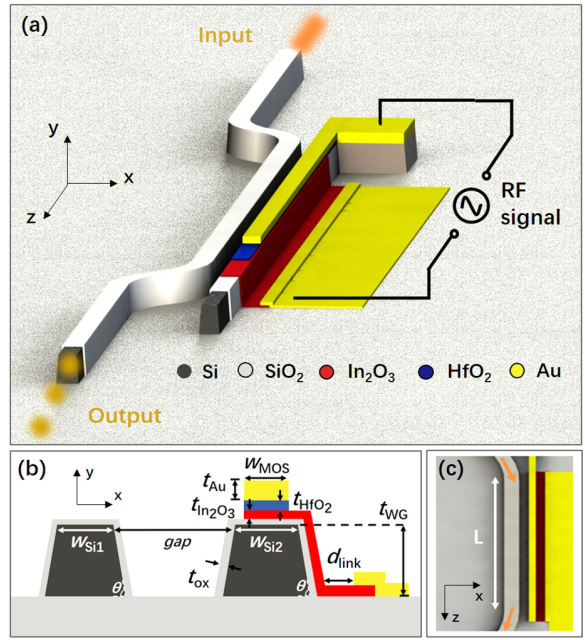

Fig. 1. (a) Schematic configuration of the proposed Si/$In_2O_3$ hybrid plasmonic modulator. (b) Cross section, and (c) top view of the asymmetric directional coupler.

Figure 1(a) shows the schematic configuration of the proposed Si/$In_2O_3$ hybrid plasmonic modulator. This modulator is realized by an asymmetric directional coupler (ADC) consisting of a silicon photonic waveguide and a Si/$In_2O_3$ hybrid plasmonic waveguide. The optical signal is modulated by radio-frequency (RF) signal applied on the Au electrodes at the top of MOS capacitor and contacting the $In_2O_3$ thin film. The cross section and top view of the ADC is illustrated in Fig. 1(b) and 1(c), respectively. All the silicon cores are covered with a silica layer. Above the silica layer of Si/$In_2O_3$ hybrid plasmonic waveguide, the MOS capacitor is made up with $In_2O_3$, $HfO_2$ and Au. The thin-film $In_2O_3$ also covers the sidewall and contacts the Au electrode on the silica substrate. With not bias voltage applied, the input light in silicon photonic waveguide is weakly coupled into the plasmonic one and the output exhibits an optical "ON" signal for its direct and low-loss transmission. When bias voltage is applied on the Au electrodes, the carrier density is changed in In$_2$O$_3$ layer and consequently so are the permittivity and refractive index of In$_2$O$_3$. The hybrid plasmonic mode is changed into an ENZ mode in the plasmonic waveguide then the phase-matching criterion is met. The input light is strongly coupled and absorbed in plasmonic waveguide then the output exhibits an optical "OFF" signal.

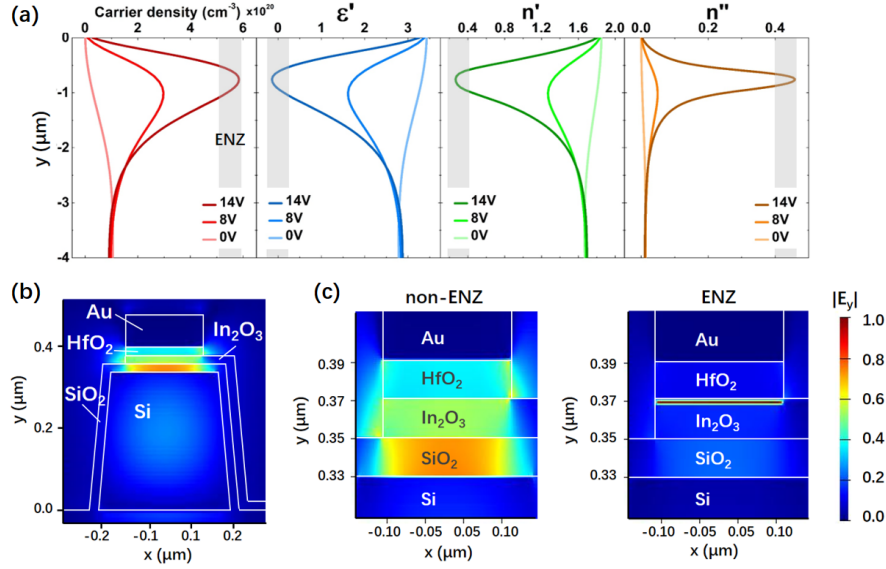

Fig. 2. (a) Calculated carrier density, real part of permittivity, real and imaginary part of refractive index in In$_2$O$_3$ with different bias voltages. (b) Simulated modal profiles (|E$_y$|) in Si/In$_2$O$_3$ hybrid plasmonic waveguide at non-ENZ state. (c) Zoomed profiles in MOS structures at both non-ENZ and ENZ states.

Figure 2(a) plots the change of carrier density distribution, permittivity and refractive index in 20nm-thick In$_2$O$_3$ under different bias, here $y$ coordinate is normalized by the interface of HfO$_2$ and In$_2$O$_3$. It is clear that when bias voltage is large as 14 V, there is a carrier density peak near the interface and in the same area the real part of permittivity is near zero, as the ENZ effect defined. The quantum moment model in Silvaco ATLAS is used to simulate carrier density distribution influenced by bias voltage, which is confirmed as the most reliable method in [13,18,24]. In our simulation here, a bias of 14 V exactly corresponds to a 5-MV/cm electric field in HfO$_2$, according to [13]. Then, Drude model is applied to describe the influence of carrier density on permittivity and refractive index for the background carrier density here is high as above $1.0\times10^{20}$ cm$^{-3}$. All the calculation above is carried out based on parameters of our In$_2$O$_3$ thin film with a mobility of 65.2 cm$^2$/V/s. With the change of refractive index of In$_2$O$_3$, the optical mode is calculated by commercial software MODE solution of Lumerical. Figure 2(b) illustrates the modal profile at non-ENZ state, and Fig. 2(c) shows profile details in MOS capacitor at non-ENZ and ENZ states. It can be seen in Fig. 2(c) that at non-ENZ state, silica, In$_2$O$_3$, and HfO$_2$ layers concentrate optical field in a decreasing account for their increasing refractive index (1.44, 1.98, 2.06), in a principle discussed in [25]. When bias voltage is applied and the carrier density near the interface of HfO$_2$ and In$_2$O$_3$ is high enough to form an ENZ area, the continuity of electric displacement vector would cause an enhancement of electric field at the interface, as the ENZ state in Fig. 2(c). From non-ENZ state to ENZ state, the optical field is turned from a low-loss plasmonic mode with thin-film confinement, into a high-loss ENZ mode with surface enhancement.

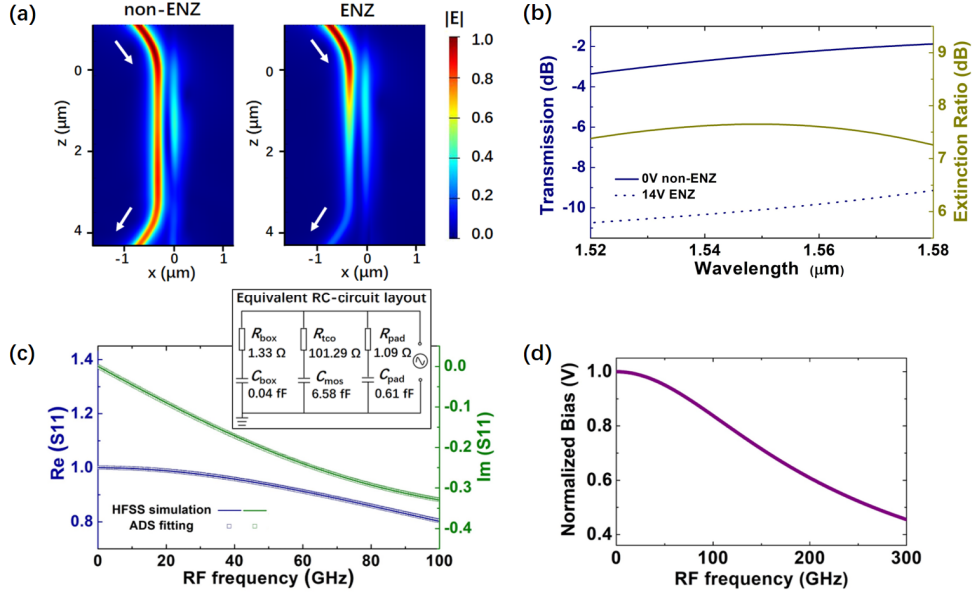

Fig. 3. (a) Simulated propagation profiles at non-ENZ and ENZ states; (b) Calculated transmission and extinction ratio spectra of the hybrid plasmonic ADC; (c) Real and imaginary parts of S11 parameters calculated from HFSS simulation and fitted with Advanced Design System (ADC), with inset of the equivalent RC-circuit layout of the $Si/In_2O_3$ hybrid plasmonic modulator; (d) Normalized voltage across the MOS capacitor as a function of RF frequency.

To utilize the strong modulation of optical field, the ADC is designed to meet phase matching criterion at ENZ state and break it at non-ENZ state, as shown in Fig. 3. Here, $t_{WG}$=340nm, $w_{Si}$ =280nm, $w_{Si2}$=220nm, $gap$=180nm, $t_{ox}$=20nm, $t_{In2O3}$=20nm, $t_{HfO2}$=20nm, and $t_{Au}$ = 80nm. To keep the resistance and optical loss in low, $d_{link}$=200nm. According to fabrication data, $\theta$ = 86°. With this design, the optical field is weakly coupled into the hybrid plasmonic waveguide at non-ENZ state and exhibits a low excess loss. While at ENZ state, the optical field is strongly coupled and absorbed as high-loss ENZ mode. Between the non-ENZ and ENZ states, the output of silicon photonic waveguide experiences a broadband modulation with a ~7.5-dB extinction ratio (ER), as shown in Fig. 3(b). The excess loss (EL) is around 3 dB in a 60-nm bandwidth. Due to the carrier transportation time in TCOs is 200~300 ps [22], the main limitation of modulation bandwidth is the resistance-capacitance (RC) delay. Based on the RC values calculated with electric-statics module in COMSOL multiphysics, we fitting the $S_{11}$ calculated from HFSS simulator with an equivalent RC circuit layout in Advanced Design System (ADS). This method has been proved to be reliable for analysis of lumped electrodes [2,18,24]. The fitted $S_{11}$ is illustrated with the calculated one from HFSS simulation in Fig. 3(c) and the inset is the result of equivalent RC-circuit layout. It can be seen that the resistance of $In_2O_3$, $R_{tco}$, is 101.29 Ω, and the capacitance of MOS, $C_{mos}$, is 6.58 fF. The low resistance results from no-doping design and the low capacitance due to the 100-nm narrow and 3.5-μm-long Au cap of MOS structure. Figure 3(d) shows that the applied bias on $HfO_2$, as a function of frequency. The RC bandwidth calculated with $f_{RC}=1/(2\pi R_{tco}C_{mos})$ is 238 GHz and can be elevated if smaller footprint can be fabricated and shorter working wavelength is permitted. It should be noted that the parasitic capacitor and resistance on electrical transmission line are not considered here, which need more optimization for an integrated chip design.

## 3. Fabrication and Measurement

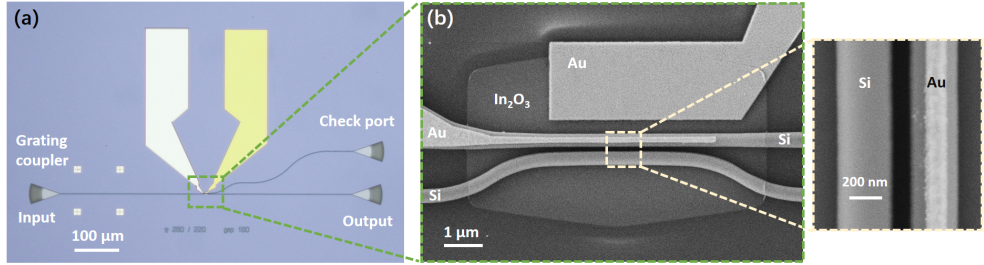

Fig. 4. (a) Optical microscope image of the fabricated plasmonic modulator; (b) SEM images of hybrid plasmonic ADC with inset of zoomed view of coupling region.

The Si/In$_2$O$_3$ hybrid plasmonic modulator was fabricated on a commercial silicon-on-insulator (SOI) wafer, with 340-nm silicon layer and 2-μm buried oxide. First, the silicon cores and grating couplers were patterned by electron-beam lithography (EBL), followed with reactive ion etching (RIE) of 340-nm depth into silicon. Then, a 2-min thermal oxidation at 1000°C was applied to form a 20-nm-thick silica layer covering silicon cores and grating couplers. Next, the In$_2$O$_3$ thin film was patterned using EBL with PMMA based photoresists and then sputtered without oxygen gas flow. After lift-off process, the film was annealed at 300°C for 1h in pressure of less than $5\times10^{-3}$ pa. The carrier concentration and mobility of the final film reach $1.14\times10^{20}$ cm$^{-3}$ and 65.2 cm$^2$/V/s, respectively, which were cross-checked by Hall measurements and ellipsometry. After In$_2$O$_3$ annealing, the negative end of electrode contacting with In$_2$O$_3$, was patterned, electron-beam evaporated with 10-nm adhesion Ti layer and 100-nm-thick Au, and lifted off. Next, a 20-nm-thick HfO$_2$ layer was conformally deposited using atomic layer deposition (ALD). Finally, the top Au layer of MOS capacitor and positive end of electrode was patterned by EBL and PMMA based photoresists, with precise alignment to the Si/In$_2$O$_3$ hybrid plasmonic waveguide. Then 3-nm adhesion Ti and 80-nm-thick Au layers were evaporated and lifted off. Fig. 4(a) shows the optical microscope image of our fabricated Si/In$_2$O$_3$ hybrid plasmonic modulator, the light-yellow area is normal Au while the dark yellow area is Au covered with HfO$_2$ layer. The grating couplers act as input and output to couple light in fiber into the chip and get the modulated optical signal. The additional check port works for fabrication quality check of ADC after first EBL and RIE processes. Fig. 4(b) shows the scanning electron microscope (SEM) image of the ADC, with different materials labeled in it. It should be added that the In$_2$O$_3$ film is sandwiched between oxidized silicon and the invisible HfO$_2$ layer, which has no boundary to be recognized by SEM. The inset of Fig. 4(b) is a zoomed view of coupling region, one can see that the silicon photonic waveguide is ~330-nm wide, he hybrid plasmonic waveguide is ~270-nm wide, the gap of ADC is ~130 nm, the Au strip on top of MOS capacitor is ~140-nm wide and the alignment mismatch of EBL was finely controlled.

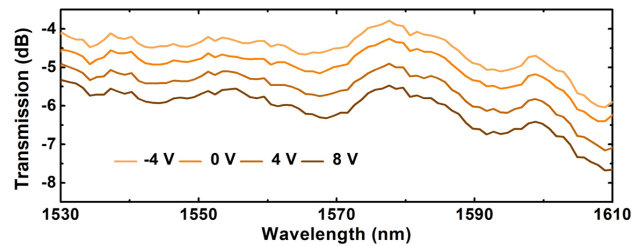

Fig. 5. Measured transmission spectra under different bias voltage, exhibiting a broadband performance.

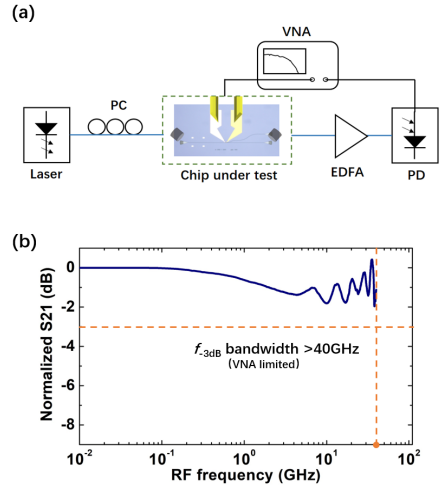

Fig. 6. (a) Measurement setup for the electro-optic frequency response, showing electrical connections (black) and optical paths (blue) using single mode fibers. (b) Normalized small-signal response $S_{21}$, exhibiting a -3dB bandwidth beyond 40 GHz.

Firstly, the broadband electrostatic modulation performance of this modulator was investigated. A broad-band amplified spontaneous emission (ASE) light source was used as the source and an optical spectrum analyzer (OSA) was applied to readout the output response. The measured results were normalized with respect to the transmission of a 500-nm-wide straight waveguide connected with grating couplers on the same chip. Figures 5(a) shows the measured transmission spectra with gradually increasing bias voltage. The fabricated modulator exhibits a 1.4-dB ER between -4 V and 8 V in 1530–1610 nm and the average excess loss (EL) is 4.5 dB. The ripple of spectra comes from fabrication difference between tested modulator and straight waveguide. One can see that the ripple slope is maintained during modulation, which guarantees the broadband performance. However, when higher bias voltage is applied and the device goes towards real ENZ region, where the ER would increase more rapidly, the $HfO_2$ layer was broken in our measurement. It shows that the fabricated $HfO_2$ film has a lower broken-down electric field than the nominal one of bulk $HfO_2$ and stable high-permittivity thin film is highly desired for modulation based on carrier dispersion effect in MOS capacitor. The small-signal frequency response ($S_{21}$) is obtained by generating a low power modulating signal (0 dBm) with a 40 GHz network analyzer, as shown in Fig. 6(a). The input light from a tunable laser with polarization controlled by a fiber polarization controller (PC) was coupled into the chip under test through a grating coupler, the input light on the chip was modulated through the RF signal and then coupled out of chip through a grating coupler again, the output light was amplified by an erbium-doped fiber amplifier (EDFA, ~25 dB) and detected by a 70-GHz photodetector (PD). It can be seen that the measured 3-dB bandwidth of the fabricated modulator is beyond 40 GHz, which is the maximal bandwidth of our VNA.

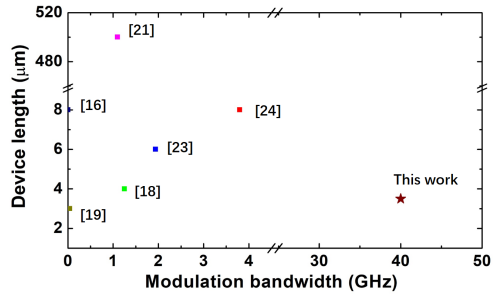

Fig. 7. Comparison with other experimental works of TCO-based modulator.

Figure 7 shows the comparison with other experimental works of our TCO-based modulator. It can be seen that modulation bandwidth is highly improved in an order of magnitude in our work. Meanwhile, the device length of our fabricated Si/In$_2$O$_3$ hybrid plasmonic modulator is relatively low compared with other works. The modulation bandwidths of devices in [16] and [19] are limited because they were pioneering works on static modulation and the high-speed design and characterization were not much considered. The device length in [21] is much larger than others for its MZI-based design, and its modulation bandwidth is limited due to its MOS capacitor cannot be narrowed for its direct integration on silicon waveguides. The modulation can be operated at GHz in [18] for its 70-Ω ion-implanted silicon, but the high capacitance from its thin and broad oxide layer of MOS capacitor hindered the elevation for higher modulation bandwidth. Based on works of [16], the doping of silicon was carefully designed and carried out in [23] in a nanocavity but the series resistance is still above 1 kΩ in that TCO-oxide-silicon MOS capacitor. The EAM in [24] has a resistance of 642 Ω due to an 8-μm resistor width of TCO and the plasmonic design. Based on that, it exhibited relatively-high 3.8-GHz modulation bandwidth. In short, the direct integrations of MOS capacitors on silicon waveguides in MZI [21], nanocavity [23], and EAMs [18,24] cannot make that the reduction of resistance or capacitance of MOS capacitor is faster than the other one. However, by reducing the length of MOS capacitor in our ADC design, the quadratically decreased capacitance and linearly increased resistance realize a decrease of RC time constant and therefore an increase of modulation bandwidth. This optimization for modulation bandwidth also intrinsically benefits for a compact footprint. Finally, as our fabricated modulator exhibits, a record-high >40 GHz modulation bandwidth is realized in a 3.5-μm device length.

## 4. Conclusion

To conclude, we designed and experimentally demonstrated a Si/In$_2$O$_3$ hybrid plasmonic modulator with a record-high modulation bandwidth beyond 40 GHz. The huge improvement of modulation bandwidth is realized by a silicon-doping-free metal-oxide-In$_2$O$_3$ capacitor integrated in a 3.5-μm-long asymmetric directional coupler (ADC). The silicon-doping-free design utilizes highly conductive In$_2$O$_3$ instead of doped silicon in MOS capacitor to lower the series resistance. Furthermore, the ADC design makes that a higher modulation bandwidth can be realized when pursuing more compact footprint. This experimental result provides a solution to break the modulation bandwidth limitation of TCO-based modulators, which is long-standing problem in the development of this field.

**Funding.** This work was supported by National Major Research and Development Program (No. 2018YFB2200200/2018YFB2200201), National Science Fund for Distinguished Young Scholars (61725503), National Natural Science Foundation of China (NSFC) (91950205, 61961146003), Zhejiang Provincial Natural Science Foundation (LZ18F050001, LD19F050001), Zhejiang Provincial Major Research and Development Program (No. 2021C01199), and the Fundamental Research Funds for the Central Universities.

**Disclosures.** The authors declare no conflicts of interest.